\documentclass[twocolumn, english, prd, nolongbibliography, superscriptaddress, aps, showpacs, showkeys, amsmath, amssymb, floatfix, nofootinbib]{revtex4-1}

\pdfoutput=1
\usepackage{graphicx}
\usepackage{color}
\usepackage[colorlinks,bookmarks]{hyperref}
\definecolor{linkblue}{rgb}{0,0,0.8}
\definecolor{linkgreen}{rgb}{0,0.5,0}
\hypersetup{linkcolor=linkblue, citecolor=linkblue, urlcolor=linkblue}

\usepackage[T1]{fontenc}
\usepackage[utf8]{inputenc}
\usepackage{lmodern}
\usepackage[USenglish]{babel}
\usepackage{enumerate}
\usepackage{nicefrac}

\newcommand{\alf}{\alpha_{\rm I} }

\begin{document}

\title{Seeding supermassive black holes with a non-vortical dark-matter subcomponent}

\author{Ignacy Sawicki}
\affiliation{Institut für Theoretische Physik, Ruprecht-Karls-Universität Heidelberg\\ Philosophenweg
16, 69120 Heidelberg, Germany}

\author{Valerio Marra}
\affiliation{Institut für Theoretische Physik, Ruprecht-Karls-Universität Heidelberg\\ Philosophenweg
16, 69120 Heidelberg, Germany}

\author{Wessel Valkenburg}
\affiliation{Instituut-Lorentz for Theoretical Physics, Universiteit Leiden\\ Niels Bohrweg 2, Leiden, NL-2333 CA, The Netherlands}

\begin{abstract}
A perfect irrotational fluid with the equation of state of dust, {\em Irrotational Dark Matter} (IDM), is incapable of virializing and instead forms a \emph{cosmoskeleton} of filaments with supermassive black holes at the joints. This stark difference from the standard cold dark matter (CDM) scenario arises because IDM must exhibit potential flow at all times, preventing shell-crossing from occurring.
This scenario is applicable to general non-oscillating scalar-field theories with a small sound speed.
Our model of combined IDM and CDM components thereby provides a solution to the problem of forming the observed billion-solar-mass black holes at redshifts of six and higher.
In particular, as a result of the reduced vortical flow, the growth of the black holes is expected to be more rapid at later times as compared to the standard scenario.
\end{abstract}


\maketitle


\section{Introduction}

In the theory of formation of cosmological large-scale structure, dark matter is modeled as a pressureless fluid---dust. However, this hydrodynamical approximation fails at shell crossing during the collapse of objects, when the velocity field becomes multivalued. From that point on, either one thinks of dark matter as particles interacting through gravity, which is the picture motivating $N$-body simulations or one has to reaverage, and translate the velocity dispersion into a pressure.

The main idea of this paper is to introduce a new subdominant component of the dark-matter sector which can be approximated to behave hydrodynamically at \emph{all relevant times}, and which \emph{always} exhibits potential flows and has a very small sound speed. We call it \emph{Irrotational Dark Matter} (IDM), as opposed to the dominant standard component, cold dark matter (CDM). A scalar field with a small sound speed and mass fulfills our requirements for IDM. Elements of such a fluid follow the geodesics of the space time at scales larger than the Jeans length and their velocity field must remain single valued and irrotational at all scales. We are not using this scalar to address the problem of dark energy, although this discussion is highly relevant in that case also.

\begin{figure}
\includegraphics[width= \columnwidth]{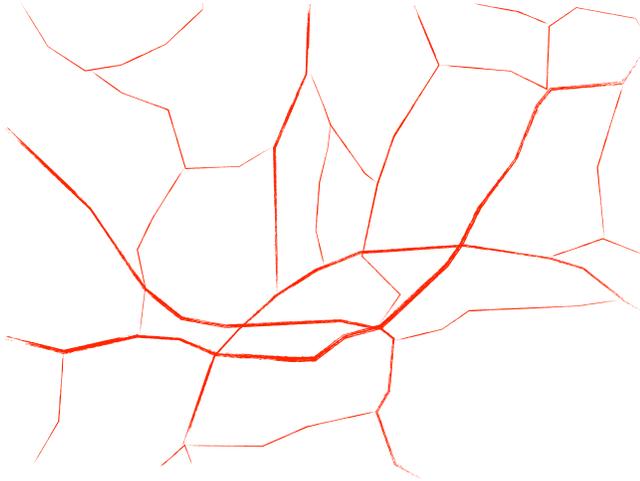}
\caption{The {\em cosmoskeleton} formed through the non-vortical collapse of \emph{Irrotational Dark Matter} in the early matter-domination era. The skeleton provides a localized potential around which cold dark matter virializes, while at its joints black hole seeds form. Picture inspired by the results of Ref.~\cite{Hidding:2012rd}.}
\label{skeleton}
\end{figure}

As long as the hydrodynamical approximation for CDM holds, the dynamics of the two dark-matter components are essentially the same. While CDM can undergo shell crossing, the trajectories of IDM approach each other until hydrodynamically supported structures are formed: planes, filaments and -- at their intersections -- approximately spherical stars, as illustrated in Fig.~\ref{skeleton}. When these stars are sufficiently compact, the pressure support becomes inadequate and they collapse to form black holes. Because of the absence of vortical flow in IDM, these structures can form very early in the history of the universe and provide a skeleton around which CDM virializes. In particular, we propose that supermassive black holes in high-redshifts quasars are seeded through the collapse of such a fluid component.

Most, if not all, galaxies have supermassive black holes (SMBHs) -- black holes of mass $M_{\rm bh} \sim 10^{6-10} M_{\odot}$ -- at their center. Our own Milky Way contains a black hole of mass $4.1 \cdot 10^{6} M_{\odot}$ \cite{Ghez:2008ms}. The Andromeda galaxy, the nearest spiral galaxy at a distance of 0.78 Mpc, contains a black hole of mass $1.4 \cdot 10^{8} M_{\odot}$ \cite{Bender:2005rq}.
However, somewhat surprisingly, SMBHs of mass $10^{9} M_{\odot}$ are observed in quasars already at redshifts $z>6$~\cite{Fan2006}, and a quasar powered by a black hole of mass $2 \cdot 10^{9} M_{\odot}$ was recently discovered at $z=7.085$~\cite{Mortlock:2011va}, when the Universe was only 0.75 Gyr old. The black holes that we observe today are presumably the dormant remnants of the powerful quasars of the past.

The physics behind the formation of these SMBHs -- a fundamental issue if one wants to understand the formation and evolution of galaxies -- is still an open problem in astrophysics.
It is not at all clear how billion-solar-mass black holes could form in less than a billion years from the Big Bang; in other words, the theory of black-hole seeds is in part still unknown or unconstrained. Various formation mechanisms have been proposed in the literature heretofore: (1) the seeds could be remnants of first-generation (Population III) massive stars which could yield black holes (BH) of mass $10-100 M_{\odot}$~\cite{Madau:2001sc}, or (2) could be the results of direct collapse of primordial gas clouds in which case seeds of mass $10^{4-6} M_{\odot}$ are produced~\cite{Begelman:2006db}. The former scenario cannot naturally reach the required mass of a billion solar masses by redshift of six, while the latter needs a very low fragmentation rate and a parent halo with a very low angular momentum.
Alternatively (3) black hole seeds could also have formed as a result of collapsing nuclear stellar clusters~\cite{Davies:2011pd}. More exotically, the seeds could be primordial, resulting directly from large quantum perturbations of curvature generated during inflation \cite{Carr:1974nx,Josan:2009qn} or even through decays of topological defects after inflation where correlated primordial-black-hole distributions can be formed \cite{Rubin:2001yw,Khlopov:2002yi,Khlopov:2004sc}. See the review papers~\cite{Volonteri:2010wz,Volonteri:2012by} and references therein for more details on the proposed formation processes.

We show that our model can produce hundreds of black hole seeds with mass $\sim10^{4}M_\odot$ in the observable Hubble volume already by redshift $z=15$. The strongest constraint arises from the relatively low mass of Milky Way's SMBH and implies that IDM can at most contribute $10^{-7}$ of the total dark-matter mass. The Milky Way's SBMH is somewhat light compared to those in neighboring galaxies, so this constraint could be weakened if one invokes a peculiar event in Milky Way's formation history.

Future probes, such as the {\em Athena+} X-ray observatory~\cite{Aird:2013xzs}, will be necessary to understand if the above mechanisms are viable or if alternative more exotic models, such as the one proposed in this paper, are required.


\section{A model for Irrotational Dark Matter}

Our idea depends on realizing a model for IDM which is hydrodynamical, not capable of producing vortical flows and which has a small sound speed and equation of state. Given these properties, the phenomenology is quite generic and does not depend on the precise model employed. Since the flows are irrotational, they are most naturally described by a scalar degree of freedom, which plays the role of the velocity potential in this picture.
 
The approach we take is to model our IDM using an action for potential-flow hydrodynamics, thinking of it as an effective action and therefore taking the dynamics as classical,\footnote{We employ the metric signature $(-+++)$ throughout and set $8\pi G_\text{N} =1=c$.}

\begin{equation}
	S_\phi = \int \mathrm{d}^4x \sqrt{-g}\, M^4 \left(\frac{X}{M^4}\right)^{\frac12\left(1+c_\text{s}^{-2}\right)} \,, \label{eq:action}
\end{equation}
where $X$ is the canonical kinetic term for the scalar $\phi$, $X\equiv-\nicefrac{1}{2}g^{\alpha\beta}\nabla_\alpha\phi\nabla_\beta\phi$, $M$ is a mass scale of this effective Lagrangian. We require the gradient of $\phi$ to be timelike everywhere, $X>0$, since only in this limit we recover a hydrodynamical description. This action falls into the k-\emph{essence} class \cite{ArmendarizPicon:2000dh,ArmendarizPicon:2000ah}, for which a formulation in terms of relativistic hydrodynamics is well known \cite{ArmendarizPicon:1999rj,Garriga:1999vw}. In this formulation, the function $\sqrt{2X}$ can be interpreted either as a chemical potential \cite{Pujolas:2011he} or as a temperature \cite{Dubovsky:2011sj}, while the parameter $c_\text{s}^2\ll 1$ represents the constant sound speed of propagation of small perturbations. Extending such actions to multiple scalar fields allows one to describe perfect-fluid hydrodynamics completely \cite{Schutz:1970my}.

The scalar field can be associated with a relativistic flow velocity,
\begin{equation}
	u_\mu \equiv -\frac{\nabla_\mu \phi}{\sqrt{2X}}\,. \label{eq:u}
\end{equation}
By virtue of Frobenius' theorem, the twist tensor for the vector field $u_\mu$ must vanish, $\perp_{[\mu}^\alpha\perp_{\nu]}^\beta \nabla_{\alpha}u_{\beta}=0$, where $\perp_{\mu\nu}\equiv g_{\mu\nu} + u_\mu u_\nu$ is the projector onto the hypersurface perpendicular to the vector $u_\mu$. This means that the flows are \emph{always} non-vortical when seen in the appropriate set of coordinates.\footnote{That is in the frame defined by $u_\mu$. In the space-time coordinates a vorticity three-vector may appear, but only inside the Jeans length, and never signifies that the fluid flows around a center. It must always be possible to slice the flow with constant $\phi$ hypersurfaces that do not cross.}

The above interpretation assigns the meaning of a clock to $\phi$. We therefore require that the scalar does not enter a period of evolution where it oscillates, since that would make this identification invalid and break down the hydrodynamical interpretation. 

The energy-momentum tensor (EMT) for this system is given by
\begin{align}
	T_{\mu\nu} &= (\rho+p) u_\mu u_\nu +p g_{\mu\nu} \,,  \label{eq:EMT}\\
	\rho &= \frac{p}{c_\text{s}^2} \,, \quad
	 p = M^4 \left(\frac{X}{M}\right)^{\frac12\left(1+c_\text{s}^{-2}\right)} \,. \notag 
\end{align}
The velocity field $u_\mu$ can now be seen to be the comoving velocity for elements of a perfect fluid. The fluid is adiabatic and has a constant equation of state $w_\text{f}=c_\text{s}^2$. We would like to stress that IDM is a dark-matter subcomponent ($w_\text{f} \ll1$) and not a dark-energy source, which we take here to be the cosmological constant.

The equation of motion for the scalar field represents the covariant conservation of a Noether charge, corresponding to the symmetry of the action \eqref{eq:action} under shifts $\phi\rightarrow\phi+\text{const}$:
\begin{equation}
	\nabla_\mu \left(X^{1/2c_\text{s}^2} u^\mu \right) = 0 \,. \label{eq:EoM}
\end{equation}
Depending on the interpretation of $\sqrt{2X}$ as either chemical potential or temperature, this Noether charge corresponds to a conserved particle number \cite{Pujolas:2011he} or conserved entropy \cite{Dubovsky:2011sj}. 

In order to ensure that the IDM collapses, forming black holes, we show in section~\ref{s:struc} that we require that the sound speed (and therefore the equation of state) be $c_\text{s}^2\lesssim 10^{-5}$. 

On the other hand, the majority of dark matter is observed to form stable halos. We parametrize the relative abundance of IDM with respect to CDM with the following constant:
\begin{equation}
\alf= \frac{\Omega_{\rm IDM}}{\Omega_{\rm DM}} \,,
\end{equation}
where $\Omega_{\rm DM} \equiv \Omega_{\rm CDM} + \Omega_{\rm IDM}$ is the total dark-matter abundance relative to the critical density in a Friedmann-Lema\^itre universe. As we will see, observations constrain $\alf \lesssim 10^{-7}$.
Given the small value of $c_\text{s}^2$, the equation of motion \eqref{eq:EoM} implies that the value of $X$ hardly changes during matter domination. The action will break down as an effective description whenever $X\sim M^4$, because at that point higher derivative terms are expected to become important. But given the upper bound on the IDM density and the extremely high exponent in the definition of $\rho$, $M$ can be easily tuned to be $M_\text{Pl}$ while still obeying the observational constraints. That is, this action gives a valid description possibly up to the Planck scale.

A simpler limit of the behavior we discuss here would be provided by an effective action for a scalar with an exactly vanishing sound speed, such as a $\lambda\phi$-fluid proposed in Ref.~\cite{Lim:2010yk}. This sort of constrained degree of freedom also appears in Ho\v{r}ava-Lifschitz theories of gravity \cite{Horava:2009uw, Blas:2009yd} where it can produce dark-matter-like dust \cite{Mukohyama:2009mz, Mukohyama:2009tp}. Axions, a class of pseudoscalar models, have long been a candidate for dark matter. However, despite their extremely small dispersion, they are massive scalar fields which oscillate during their evolution and therefore their value cannot be used as an affine parameter along trajectories, as we need to do in our scenario \cite{Sikivie:2006ni,Jaeckel:2010ni}.

An alternative to the effective-action picture above is to consider quantum behavior of Bose-Einstein condensates (BEC). One can describe the dynamics using the Gross-Pitaevskii equation for the wavefunction of the BEC and then rewrite that as an Euler equation for a fluid with the phase of the wavefunction playing the role of the velocity potential. One then finds that two effective pressure terms appear with positive contributions: quantum pressure, i.e.\ resulting from the de Broglie wavelength of the condensate, and a pressure from any self-interaction terms. The former appears when the mass of the condensate is small enough, such the condensate cannot be localized as a result of the uncertainty principle. This was exploited in the Fuzzy CDM model \cite{Hu:2000ke} to erase the CDM power spectrum at small scales, removing cusps at the centers of DM halos. Self-interaction terms change the pressure at high densities, i.e.\ make the sound speed density dependent which can then support alternative static solutions, e.g.~\cite{Peebles:2000yy}.\footnote{This scenario is different to the self-interacting CDM model of Ref.~\cite{Spergel:1999mh} where standard particle DM has the usual weakly interacting massive particle cross-section, but is not in a condensed state and therefore can produce the usual vortical flows of CDM. Nonetheless, such models also enhance BH growth \cite{Hennawi:2001be}.} Interestingly, for our model to work in the BEC scenario, we require the opposite limit for both of these features: small de Broglie wavelengths/larger boson masses so that the BEC can collapse on small scales and a tuned down self-interaction term. This limit has the added feature of preventing any vortices from forming in the condensate \cite{RindlerDaller:2011kx}, maintaining the irrotational fluid property that we require. The question remains whether is it possible to both have a large-enough mass to reduce quantum pressure sufficiently and to still allow for condensation in the first place. 

Interestingly, if axions thermalize, they would form a BEC, although not in the parameter range that we require: they would be capable of supporting stable structures \cite{Sikivie:2009qn}.

We would also like to point out that our setup does not have to arise from a degree of freedom completely separate from CDM: it might be possible for cold dark matter to form a scalar condensate and not populate it fully as a result of, e.g.\ too high a temperature. This sort of mechanism might be able to naturally produce the sort of hierarchy between the CDM and the IDM densities that we require.

Dark-energy models featuring a non-oscillating scalar field with a small sound speed would exhibit similar flow properties when non-linear. However, in most such perfect-fluid setups, the growth rate is very low as a result of the value of the equation of state. Spherical collapse in such models was discussed in Ref.~\cite{Creminelli:2009mu}. However, if perturbations in the ghost condensate model \cite{ArkaniHamed:2003uy, ArkaniHamed:2005gu} ever become non-linear, they will behave in a way that is very similar to the sort of phenomenology we describe in this paper. Generalised Chaplygin gas \cite{Kamenshchik:2001cp, Bento:2002ps} and  unified dark-matter models \cite{Scherrer:2004au, Bertacca:2007ux, Gao:2009me} have an equation of state that is very close to dust until late times and therefore yet again the physics described in this paper would be realized in these models at least until the onset of acceleration.


\section{Structures large and small in IDM}
\label{s:struc}

\begin{figure*}
\includegraphics[width=.8 \columnwidth]{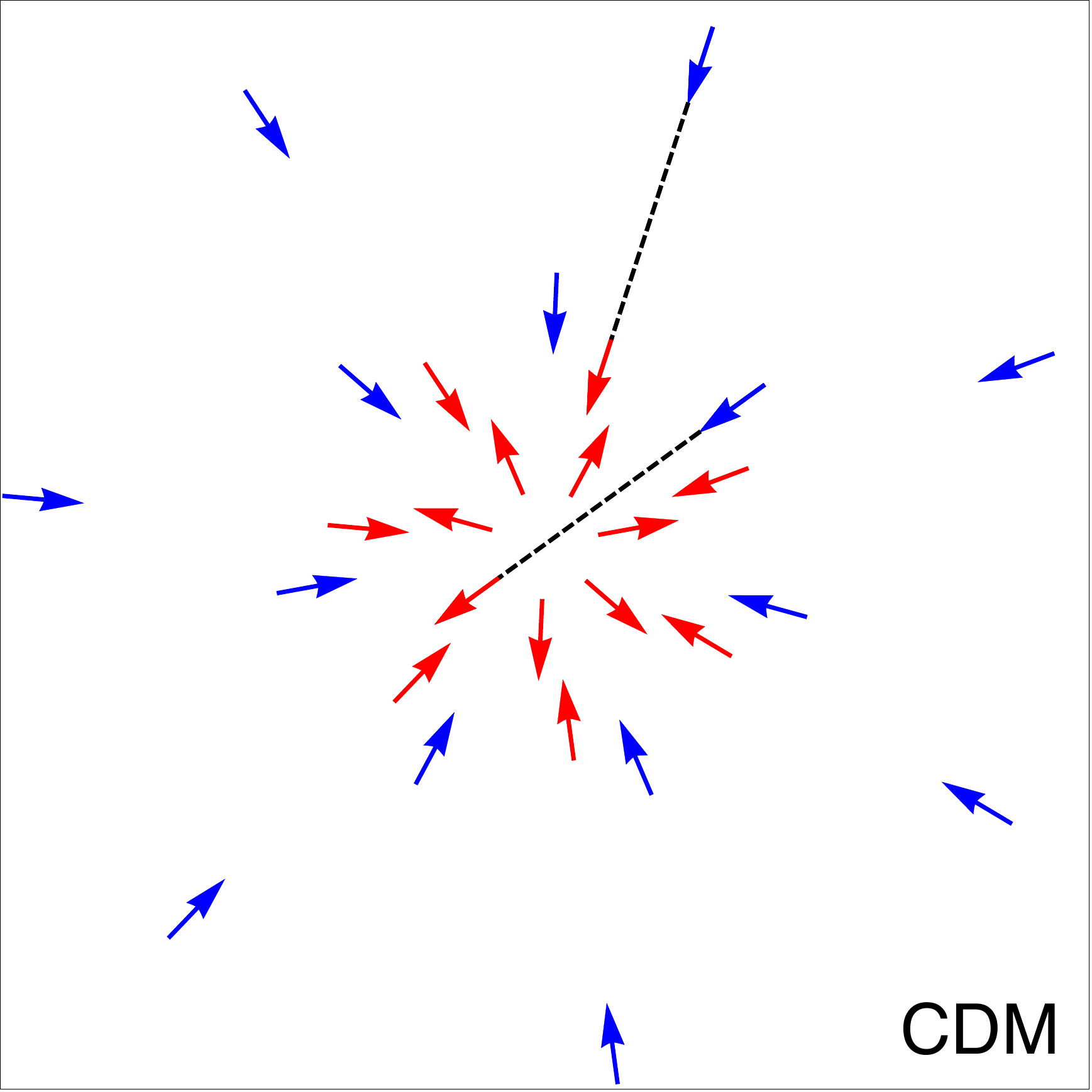}
\qquad \qquad
\includegraphics[width=.8 \columnwidth]{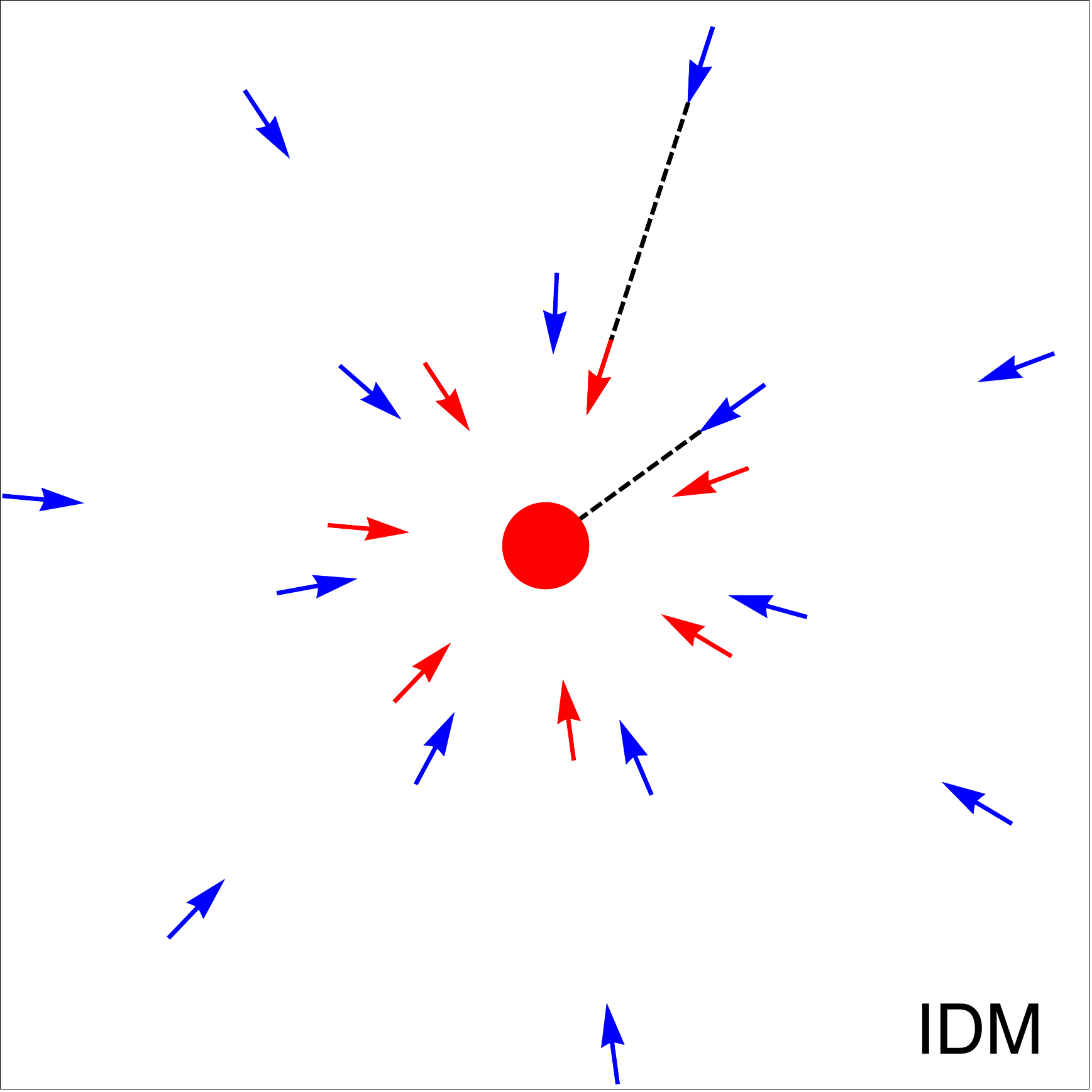}
\caption{
A slice through a spherically symmetric configuration undergoing collapse.
The arrows depict the velocity of particles (in the case of CDM, left panel) and fluid elements (in the case of IDM, right panel).
Two time steps are represented in the sketch: blue represents a time $t_{1}$, red a later time $t_{2}>t_{1}$.
The dashed line helps to guide the eye so as to track the movement of the arrows between the two time steps.
This picture shows how the collapse of CDM and IDM is the same up to shell crossing at the center: in the CDM case the particles pass through each other effectively unhindered, while in the IDM case the fluid elements cannot cross and coalescence into central accreting structure.
}
\label{IDMvsCDM}
\end{figure*}

Let us now discuss cosmological structure formation in the IDM fluid, contrasting it with the $N$-body picture pertinent to CDM.

We assume that density perturbations in the two fluids are adiabatic, i.e.\ that the density contrast in both DM components is effectively the same, since $w_\text{f}\ll 1$. If the two components are related then this is a natural condition. Otherwise, this requires both subcomponents to have been produced from the decay of the inflaton or at least to have been in thermal equilibrium at some point in the early universe. Strictly speaking, if \eqref{eq:action} described IDM also during inflation, it is a light spectator scalar field and it gets its own perturbations, which would be an isocurvature mode. However, since the equation of state is nearly that of dust, the isocurvature mode will decay during subsequent evolution \cite{Malquarti:2002iu}.

Since IDM is a scalar, it cannot carry vector and tensor perturbations. As these are subdominant in inflationary initial conditions and have no sources up to quadratic order, they are not relevant for the CDM density distribution either. Since $w_\text{f}=c_\text{s}^2\ll 1$, the evolution of both the components on background and linear-perturbation level will be essentially the same. 

The sound speed $c_\text{s}^2$ provides a Jeans length for the perturbations of IDM. On scales $c_\text{s} k/aH\gg 1$, the scalar linear perturbations receive pressure support and therefore oscillate in momentum space. However, our requirement \eqref{cs2} means that this scale would be irrelevant for the observed clustering, whatever the value of $\alf$ \cite{Sandvik:2002jz}.\footnote{Strictly speaking, the measured power spectrum is of galaxies which are biased with respect to the gravitational potential. Such oscillations in the potential may still be allowed by data if galaxy bias were related to the CDM component and not the total density perturbation. See Refs~\cite{Amendola:2012ky,Motta:2013cwa} for a discussion of what is actually observable in late-time cosmology.}

The difference between the two DM subcomponents arises at the moment the CDM flow undergoes shell-crossing, only deep within the non-linear regime. It is very stark. 
CDM particles collide and pass through each other unhindered, hardly changing their velocities, unless they happen to have directly interacted. The representation of CDM as a pressureless dust breaks down at this very moment since velocities become properties of the particles and no longer a single-valued vector field: CDM is no longer a pressureless fluid after shell crossing. One can re-average over volume elements and reinterpret the CDM as a fluid with a non-zero pressure representing the internal velocities of the particles. This fluid can then potentially form static solutions: halos. Given a large-enough localized overdensity, the CDM will recollapse and form a bound structure, eventually virializing through gravitational interactions. The evolution between shell crossing and the final static halo necessarily experiences a period where the vorticity of the averaged velocity field is large compared to the divergence of the velocity \cite{Pueblas:2008uv}.

The IDM velocity field $u^\mu$ is derived from a scalar field and therefore the flow must be irrotational and single-valued at all times where the effective action \eqref{eq:action} is valid. This is in direct opposition to the CDM case. Let us for the moment discuss the limit of $c_\text{s}^2=0$. We should point out that in this limit the evolution forms caustics, submanifolds on which gradients are divergent since the scalar-field value depends on the direction of approach. We later reintroduce a small sound speed, which will not change the general behavior, but can  resolve the caustics at least under some circumstances, e.g.\ in the spherically symmetrical case of interest here.\footnote{See Ref.~\cite{Felder:2002sv} where caustics still appear with a non-zero sound speed. Caustics in field theories with non-linear kinetic terms may be inevitable, but since they involve divergent gradients, they correspond to singular energy densities and therefore defects such as domain walls or cosmic strings, or indeed black holes.}

In IDM with a vanishing sound speed, the fluid elements \emph{always} follow geodesics of the space time (see e.g. \cite{Lim:2010yk} or the geodesic choice for the models of \cite{DeSantiago:2012xh,Wang:2013qy} for such a model). Therefore the infall is the same as for the CDM initially. However, at the same moment that the CDM undergoes shell crossing, the zero-sound-speed fluid forms a caustic: a singular manifold where many trajectories coalesce. This is the only solution possible if the velocity field is to remain irrotational and single-valued and yet there is no pressure support. These caustics can be of any dimension, depending on the initial configuration: walls, filaments, and at their intersections -- point-like singularities, and form a structure we call the \emph{cosmoskeleton}, see Fig.~\ref{skeleton}. 

In the limit of zero sound speed and no gravity, the IDM picture becomes identical to the limit of zero viscosity in the adhesion model~\cite{Gurbatov:1989az,Weinberg:1990ej}: the adhesion model extends the Zel'dovich approximation of structure formation in which elements are taken to have constant velocities and pass through each other without interaction to a model where the elements slow each other down through viscosity, or even stick together.  A good illustration of this picture is provided by the simulations of Ref.~\cite{Hidding:2012rd} where the adhesion model is applied to CDM.

In our scenario, there are two components, IDM and CDM, where the IDM behaves similarly to the zero-viscosity limit of the adhesion model, while CDM gains a multi-valued velocity field and virializes. 
We illustrate the difference between the two subcomponents in a slice through a spherically symmetric configuration in Fig.~\ref{IDMvsCDM}.

When gravity is included, the caustics become singular gravitational solutions: filaments are cosmic strings while at their intersections black holes are formed. These objects continue to accrete the IDM subcomponent. However, this occurs relatively rapidly since the scalar is not capable of virializing as a result of being able neither to have a vortical velocity field nor to provide pressure support.  Thus the picture of structure formation is altered: first, the irrotational fluid collapses to form a \emph{cosmoskeleton}, illustrated in Fig.~\ref{skeleton}, with its topological structure determined by the initial power spectrum of perturbations. The CDM then virializes around this structure, in the presence of localized gravitational potentials.

When a sufficiently small sound speed is reintroduced, the gross picture does not change. The irrotational infall of IDM still proceeds to form the \emph{cosmoskeleton}. However, depending on the initial conditions, the infall might get arrested by the pressure and a static object might be formed.\footnote{We should point out that these static solutions are hydrodynamical, i.e.\ have $\partial_\mu \phi$ timelike everywhere. What is spacelike is the gradient of $X$, which provides the description of the density field as a function of radius. These solutions are of the type described in Refs~\cite{Akhoury:2008nn} or \cite[section 4]{Pujolas:2011he}, rather than the spacelike $\partial_\mu\phi$ solutions investigated in Refs~\cite{ArmendarizPicon:2005nz, Diez-Tejedor:2013sza}.}. 

In particular, fluids with a constant sound speed have the singular isothermal sphere (SIS) as their spherical static solution. 
The density for this solution behaves as $r^{-2}$, while the mass inside a radius $r$ is 
\begin{equation}
	M_\text{SIS}(r)=16\pi c_\text{s}^2 r \,,
\end{equation}
(see e.g.\ \cite[pg.\ 305]{Binney2008}). If the total mass $M$ of a particular spherically-symmetric initial configuration of radius $r_\text{i}$ satisfies $M>M_\text{SIS}(r_\text{i})$, then the pressure support that IDM can develop will never be sufficient to arrest the collapse that ensues. This can be re-expressed as a condition on the gravitational potential difference between the center and the edge of this overdensity. If
\begin{equation}
	\left|\Phi(r_\text{i})-\Phi(0)\right| \gtrsim 2c_\text{s}^2
\end{equation}
then the IDM overdensity will never be able to form an SIS and will collapse to a black hole. Since inflation sets up nearly scale-invariant perturbations with normalization approximately $\sim10^{-5}$,  we obtain the condition 
\begin{equation}
	c_\text{s}^2\lesssim |\Phi| \simeq 10^{-5} \,. \label{cs2}
\end{equation}
which ensures that essentially all primordial roughly spherical IDM overdensities produce black holes.

Moreover, since IDM only makes up a small part of the total DM density, $\alf\ll 1$, the gravitational potential is dominated by the evolution of dark matter. Therefore even if the initial gravitational potential is small enough to allow IDM to form a static solution, a subsequent formation of a bound CDM halo will likely deepen the local potential well and trigger a collapse of the static IDM configuration. A generic prediction of our model is that any bound DM object where the velocity dispersion $\sigma_{v}^2>2c_\text{s}^2$  has -- or at least has had -- a collapsed IDM black hole at some point (see section~\ref{s:abund} for an explanation of the ambiguity).


\section{Black-hole formation and abundance}
\label{s:abund}

\begin{figure}
\includegraphics[width= .93 \columnwidth]{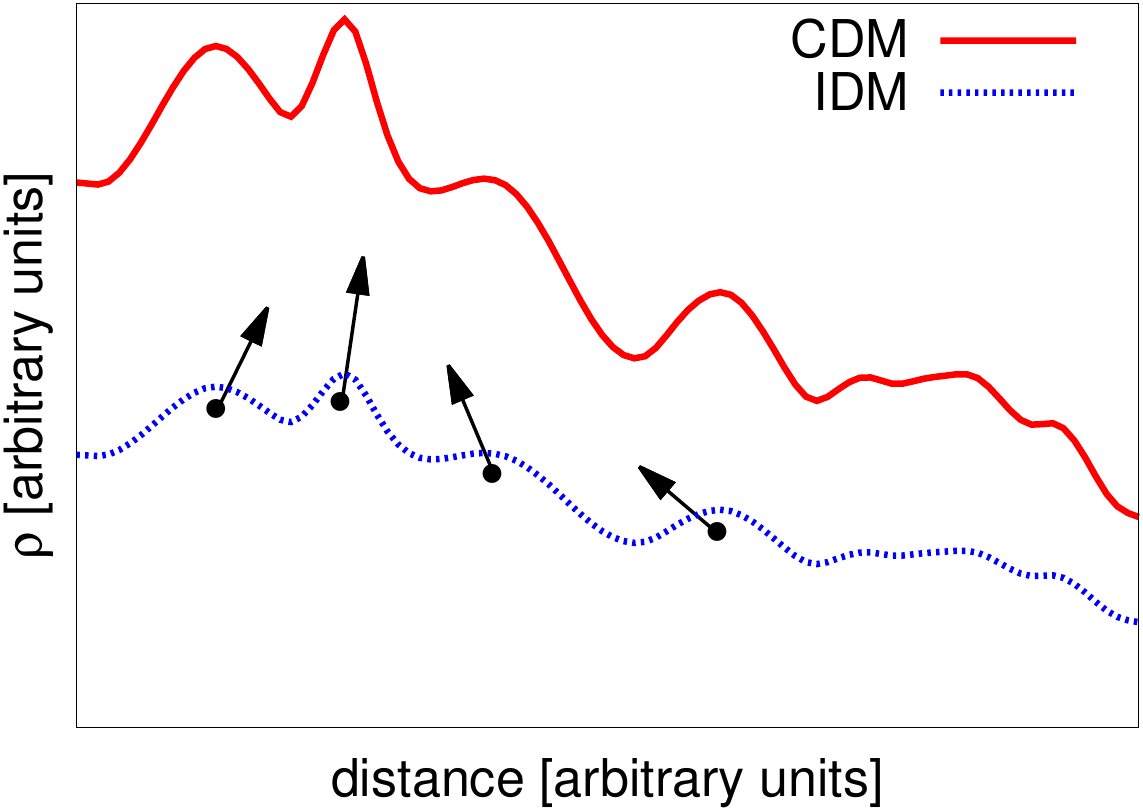}
\caption{The scenario proposed in this paper is one where black holes form in the IDM component at the center of any collapsing structure, and light black holes continue to merge into a heavier one in a larger-scale over-density. The \emph{Irrotational Dark Matter} is subdominant with respect to the cold dark matter.}
\label{sketch}
\end{figure}

We now turn to a discussion of the abundance of such black hole seeds formed from IDM. We assume adiabatic initial conditions, as illustrated in Fig.~\ref{sketch}. Given any collapsing spherical overdensity, our model predicts that the black hole seeds are formed from the IDM approximately at the moment of CDM shell crossing and therefore prior to the halo's virialization provided that the sound speed satisfies condition \eqref{cs2}. This means that a black hole seed forms at the center of \emph{any} structure that will eventually virialize. Since IDM is irrotational, these black-hole seeds would not carry angular momentum initially. Over time, the seeds accrete baryons and CDM in the standard way, and therefore any angular momentum they may be carrying and thus would be generically expected to spin up.

The overall picture of structure formation is the standard hierarchical one: structures on smaller scales would collapse first (smaller peaks in Fig.~\ref{sketch}) and then proceed to collapse into larger structures at later times (overall overdensity in Fig.~\ref{sketch}). This scenario implies that \emph{every} collapsed structure of mass $M_{\rm halo}$ initially contains a black-hole seed of mass $\alf M_{\rm halo}$. 

Given a history of mergers that large halos undergo over their evolution, the mass of the IDM seed at the center of a halo of mass  $M_{\rm halo}$ would be
\begin{equation}
	M_{\rm seed} = \alf F \, M_{\rm halo}\,, \label{eq:ms}
\end{equation}
where the quantity $F\sim 1$ parametrizes the theoretical uncertainty on details of the mass accretion history of a halo and of the IDM accretion itself which may depend on the precise value of the IDM sound speed. 

\begin{figure}
\includegraphics[width=\columnwidth]{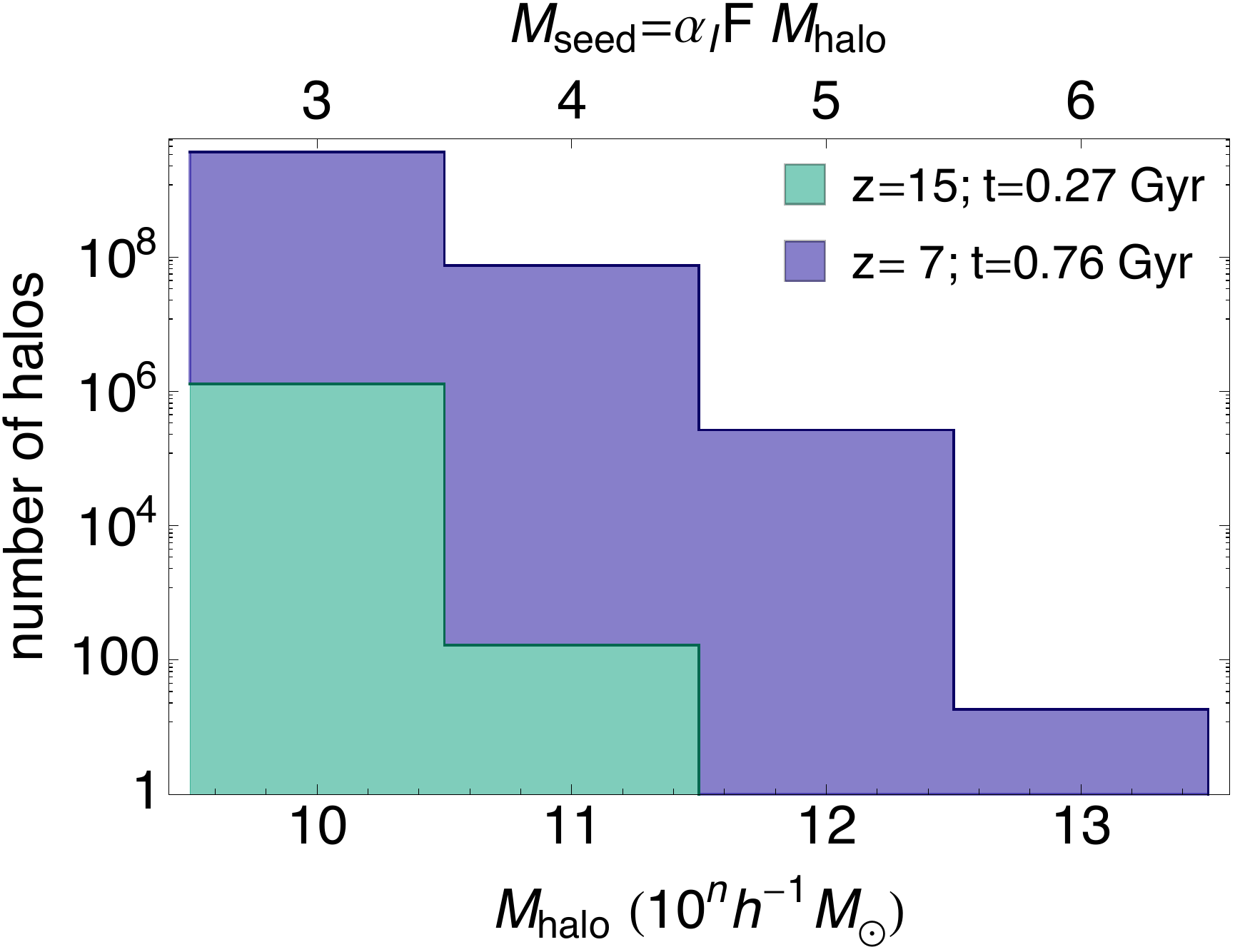}
\caption{Abundance of halos (mass on bottom horizontal axis) and black hole seeds (mass on top horizontal axis) in a Hubble volume $(c/H_{0})^{3}$ for $\alf F= 10^{-7}$ for various values of redshift and mass. By redshift $z=15$ we expect hundreds of black-hole seeds of mass $M\sim10^{3-4}M_\odot$ in the observable volume of the universe. Just the subsequent accretion of baryons is enough~\cite{Volonteri:2010wz} to bring such a black-hole seed to the mass of $10^9 M_\odot$ observed in Ref.~\cite{Mortlock:2011va} by redshift $z=7$. In our scenario, the BH seeds continue to accrete IDM from the filaments of the \emph{cosmoskeleton} and therefore will grow even faster, in principle by a couple of orders of magnitude between these two redshifts, in line with the growth of their parent halos. For this plot we have used as fiducial cosmology the latest Planck+BAO constraints~\cite{Ade:2013zuv}.
See Section \ref{s:abund} for more details.} 
\label{hab}
\end{figure}

The Milky Way has an estimated mass (including DM) of $10^{12} M_{\odot}$~\cite{McMillan:2011wd} and its SMBH an estimated mass of $4 \cdot 10^{6} M_{\odot}$, which is notably a significantly smaller proportion of the halo mass than SBMHs at the center of nearby galaxies (e.g.~Andromeda). This puts the strongest constraint on the product $F\alf \lesssim 10^{-7}$: the IDM black hole seed needs to be much lighter than the present-day central black hole, which has been accreting mass from its surrounding environment.  $N$-body simulations suggest that halos with the size of the Milky Way could have grown substantially during their evolution history~\cite{Wechsler:2001cs}.
Assuming conservatively $F\lesssim1$, this implies that $\alf = 10^{-7}$ satisfies the constraint above.

Before proceeding, it is important to comment on the evolution of the black-hole seeds.
In the standard scenarios, one wishes to have massive ($\gtrsim 10^{3} M_{\odot}$) seeds in protogalaxies of $\sim10^{8} M_{\odot}$. The seeds then grow by accretion of baryons and mergers into the billion-solar-mass black holes observed at $z=7$. This process can be inefficient: (1) three-body interactions between black-hole seeds frequently result in the third BH being ejected through a sling shot; (2) BHs which have been spun up by infalling matter do not accrete further very efficiently~\cite{1964ApJ...140..796S,Sesana:2013dma}.

In the case of IDM, the initial seeds in the protogalaxies would be very light. 
However, as hierarchical structure formation goes on, the smaller seeds are expected to keep merging very efficiently. One has to remember that our black-hole seeds are not yet static Schwarzschild/Kerr solutions, but they are still accreting the IDM from the filaments at the intersection of which they lie. These black hole seeds would not yet have shed their hair and their flow is still restricted to be irrotational. This means that the black-hole seeds sling-shots should not happen. The IDM black-hole seeds are much more likely to merge than in the standard scenarios. As a consequence, $F$ should at any time be not significantly different from unity and the BH seed mass should grow in line with the parent halo's at all times.

Another consequence of the presence of BH hair is that whenever there is large relative angular momentum locally in the CDM flow during mergers, the IDM will not be able to move with it and therefore the BHs are likely to be offset from centers of halos. 
The offset must be at intra-halo scales, as at larger scales positions of CDM over-densities are correctly described by the (truncated) Zel'dovich Approximation~\cite{1993MNRAS.260..765C}. That is, a non-vortical setup predicts structures at the same locations as a full N-body simulation of pure CDM on larger scales. As a consequence, on such large scales the IDM structure traces exactly the CDM structure, and only on smaller scales, intra-halo, one can expect the positions of IDM structure to start deviating from the positions of CDM structure.

A careful numerical analysis will be necessary in order to quantify the statistics of offsets of central black holes from galactic centers. This is could be studied using the adhesion model (see e.g.~\cite{Hidding:2012rd}) endowed with baryonic physics at the relevant sub-galactic scales, something that has not yet been explored.

In the circumstance of a merger with a very large angular momentum, it may turn out that the IDM cannot follow the CDM flow at all and the black hole might be ripped out of the parent halo, thus producing BH-less galaxies.
Indeed an example of this process having occurred could be M33: this galaxy of halo mass $M\sim 10^{10-11} M_\odot$ is very unusual in having no bulge \cite{Corbelli:2003sn}. Its central black hole, if it exists,  is constrained to have $M<3000 M_\odot$ \cite{Merritt:2001rd}. In principle, the Milky Way constraint above is just compatible with this case. However, the lack of bulge may signify that the black hole was ripped out in the past and the standard BH-bulge feedback never took place. There are other examples of such galaxies, see e.g.~Ref.~\cite{Kormendy:2010nc}, although the constraints on the BH masses are much weaker.
Galaxies without BH, either satellite or in a group, clearly have $F=0$. However, at the scale of the corresponding parent halo $F$ should still be not far from unity, as argued above. Again, numerical studies would be be necessary to precisely quantify the statistics of the $F$ parameter.
\newline

To predict the abundance of black-hole seeds, we can use existing mass functions, since IDM and CDM collapse at linear level along effectively the same geodesics. As we have already explained, any relaxed halo with velocity dispersion $\sigma^2_v>2c_\text{s}^2$ contains an IDM seed black hole (since $\sigma_{v}^{2}$ is equal to the local Newtonian potential). We show in Fig.~\ref{hab} the number of halos as a function of mass and redshift in a present-day Hubble volume $(c/H_{0})^{3}$ which should roughly represent the observable universe.
We have used the mass function of Ref.~\cite{Warren:2005ey}, which allows us to compute halo abundances at high redshift.\footnote{However, our results do not strongly depend on the mass function adopted. For example, if the mass function of Ref.~\cite{Courtin:2010gx} is used instead, roughly twice as many halos in the relevant mass range are predicted.}
From the plot we can conclude that already at a redshift of $z=15$ there should be hundreds of halos with mass in the range $10^{10-11} h^{-1} M_{\odot}$, each one with a seed black hole.
If we take $F=1$ (the value of $F$ depends on the mass accretion history up to the given redshift) the seed black holes have the mass of $10^{3-4} h^{-1}  M_{\odot}$.
According to Ref.~\cite{Volonteri:2010wz}, given an initial black hole seed in this mass range, standard accretion of baryons would bring the mass to the desired $10^9 M_\odot$ within 0.5 Gyr. This happens to be equivalent to redshift $z=7$ and therefore would easily allow for the existence of the SMBH of Ref.~\cite{Mortlock:2011va}.

We stress again that the masses of our black-hole seeds continuously grow as a result of constant IDM accretion from filaments on top of the standard CDM/baryon accretion, in line with the growth of the parent halos.
As halos with mass in the range $10^{10-11} h^{-1} M_{\odot}$ are expected to grow by as much as two orders of magnitude between redshift $z=15$ and $z=7$, a similar IDM accretion is predicted for their black holes.
Consequently, the CDM/baryon accretion discussed above does not have to explain the black hole growth alone and can be possibly modeled more conservatively.
Moreover, we would expect the low angular momentum of these supermassive black holes to increase the accretion rate as it ensures a lower mass-radiation conversion efficiency~\cite{Sesana:2013dma}.
In conclusion, the model presented in this paper succeeds in predicting the observed supermassive black holes.

\section{Potential observational signatures}

The scenario proposed in this paper predicts a wealth of new physical effects which may be used to falsify it given numerical studies to understand the detailed statistics in our scenario. We have grouped the observational signals into two categories, one regarding the evolution of black holes and their host galaxies, the second -- regarding a possible impact of this scenario on cosmological observations.

\subsection*{Evolution of black holes and parent galaxies}

\begin{enumerate}[(i)]
\item Black holes inside dwarf galaxies should form, provided that the dispersion is higher than the IDM sound speed and the back hole is not ripped out as a consequence of mergers (see (iii)). This has already been observed in Ref.~\cite{Reines:2011na}. Within the standard $\Lambda$CDM model, dwarf galaxies should not have massive accreting black holes due to the poverty of baryons.

\item IDM black holes continue to accrete the IDM from the filaments in which they lie. As long as this continues, the BHs will be attached to the \emph{cosmoskeleton} and move with it according to potential flow: they are not stationary BH solutions but rather have hair. The dynamics of CDM is much less restricted, especially during mergers, which might result in the IDM BHs lying off-center in the halos, depending on the evolution history of the particular halo. In that sense this scenario is appropriate for the test of the equivalence principle proposed in Ref.~\cite{Hui:2012jb}.

\item An interaction between halos which have very high relative angular momentum could lead to an IDM black hole being ripped out of its parent halo, since IDM cannot follow the vortical flow.
BH-less galaxies are therefore predicted.
If bulges are a product of BH feedback, this mechanism could explain bulge-less galaxies such as the M33 \cite{Merritt:2001rd}.

\item Supermassive black hole properties are correlated with the host galaxy properties~\cite{Volonteri:2010wz}. This signals an important feedback between black hole and galaxy formations. This feedback could be different within our scenario as the growth of the SMBH is expected to be faster at later times than in the standard scenario because of the reduced vortical flow~\cite{Choi:2013kia}.

\item IDM black-hole seeds do not carry angular momentum initially. Over time, the seeds are expected to spin up as they accrete baryons and CDM, and therefore accrete any angular momentum the baryons and CDM carry.
The mismatch in the spin of observed supermassive black holes compared to other formation history could then either confirm or falsify the scenario proposed in this paper~\cite{Reynolds:2013rva}.

\end{enumerate}

As discussed in the previous Section, a quantitative comparison with present-day and forecasted observations would be possible only after careful numerical studies of galaxy evolution in the presence of an IDM component.
A possible route -- not yet taken -- could be based on endowing the adhesion model with baryonic effects.
Alternatively, one has to directly solve the full dynamical scalar field equations of motion in a CDM N-body simulation, as done for instance in Ref.~\cite{Llinares:2013qbh} (see also Ref.~\cite{Christopherson:2012kw}).

\subsection*{Cosmological observations}

The difference in the clustering of IDM and CDM at non-linear scales would in principle affect cosmological observables. This effect is always suppressed by $\alf$ and therefore would only be able to provide significant constraints for larger values of this parameter.

\begin{enumerate}[(i)]

\item The bones of the \emph{cosmoskeleton} would be similar to cosmic strings, especially in the limit of vanishing sound speed. They would form during matter domination and therefore not have an impact on the formation of the cosmic microwave background (CMB). However, they would lens the CMB at later times, inducing non-Gaussianities in the observed maps. One can roughly use the Planck constraints on cosmic-string tension $G\mu<10^{-6}$ \cite{Ade:2013xla} and compare this with the mass per unit length of the expected CDM filaments in $\Lambda$CDM of $\mu\sim 10^{6-7} M_\odot/\text{pc}$, as can be estimated from the simulations of Ref.~\cite{Colberg:2004cd}. These estimates imply that such cosmic-string signatures give the rather weak constraint $\alf\lesssim 1$.

\item The IDM structure formation is faster than the standard CDM one. Therefore one expects an altered Rees-Sciama effect on the CMB, depending on the magnitude of the IDM sound speed and $\alf$. A careful simulation of IDM would be necessary to understand how much stronger the effect is, but it would clearly by suppressed by $\alf$. A combination of Planck and LSST weak-lensing data is expected to detect the Rees-Sciama effect only at $1.5\sigma$ and therefore is unlikely to be constraining for our scenario \cite{Nishizawa:2007pv}. However, for cosmic-variance-limited surveys this improves to $50\sigma$ and therefore should provide stronger constraints than (i).

\end{enumerate}

\section{Conclusions}

We have proposed a new dark matter subcomponent, {\em Irrotational Dark Matter}, which behaves like cold dark matter at background and linear level but, as a consequence of obeying irrotational hydrodynamics at all times, cannot shell cross and therefore forms a \emph{cosmoskeleton} of IDM structures in the early matter era. At the joints of this \emph{cosmoskellington} black-hole seeds grow.
The scenario proposed in this paper predicts many new physical effects which may in principle be strong enough  to falsify it.

This paper shows that models which are fundamentally irrotational (scalar-field theories) have very different non-linear structure-formation scenarios. Dark energy is usually considered as non-clustering. But more exotic models, which can have large perturbations, should exhibit some of the behavior discussed here (see e.g.~the dark-energy web of Ref.~\cite{Gerasimos2013}), although the low equation of state hinders clustering~\cite{Creminelli:2009mu} in perfect-fluid models. Since dark energy dominates the matter density at late times, this sort of \emph{cosmoskeleton} scenario could be realized and gravitationally significant, allowing for constraints to be put on large classes of clustering dark energy. 

Moreover, generalized Chaplygin gas or unified-dark-matter models featuring a single scalar degree of freedom for both dark matter and dark energy will not form vortical flows. Since all the dark matter is made up of the scalar in these models, if the sound speed is sufficiently small (as it must be to evade the constraints of Ref.~\cite{Sandvik:2002jz}) and the scalar dos not oscillate, it is predicted by our discussion to irrotationally collapse into black holes instead of forming the observed stable halos. This puts significant pressure on the viability of these scenarios in general.

As we have mentioned earlier, one could imagine that a partial condensation of CDM could provide a model for IDM. Such a model could potentially explain the smallness of $\alf$ and would be a most natural mechanism for the production of the irrotational component. Whether such a realization is at all possible, and whether it would be stable in the presence of the large density gradients and tidal forces at centers of collapse is an important avenue for investigation.

In terms of the structure formation in this mixed CDM/IDM scenario, the next step -- necessary if one wants to conclude whether this model can effectively produce black hole seeds of the desired parameters  -- is numerical modeling. It must be stressed that one {\em cannot} model the IDM subcomponent by means of particles as sometimes is done in $N$-body simulations, but one directly has to solve the scalar field equations of motion. This observation applies also to scalar-field models of dark-energy.

\
\begin{acknowledgments}

It is a pleasure to thank L.~Amendola, M.~Amin, M.~Baldi, A.~Bueno Belloso, P.S.~Corasaniti, C.~Germani, J.~Hennawi, M.~Kunz, E.A.~Lim, A.~Macci\`{o}, N.~Nunes, G.~Rigopoulos, A.~Rodigast, I.~Saltas, M.~Scherer, A.~Vikman and D.~Wands for illuminating discussions.
V.M.\ and I.S.\ acknowledge funding from DFG through the project TRR33 ``The Dark Universe''. W.V.\ is supported by a Veni research grant from the Netherlands Organization for Scientific Research (NWO).

\end{acknowledgments}

\bibliographystyle{utphys}
\bibliography{refs}

\end{document}